\documentclass[times, 10pt,twocolumn]{article}
\usepackage{latex8}
\usepackage{times}
\usepackage{amsmath}
\usepackage{amssymb}

\begin{document}

\title{Managing Delegation in Access Control Models}

%
\author{Meriam Ben Ghorbel-Talbi$^{a,b}$, Frederic Cuppens$^a$, Nora Cuppens-Boulahia$^a$, Adel Bouhoula$^b$ \\
$^a$GET/ENST Bretagne,\\ 2 rue de la Chataigneraie, 35512,
Cesson Sevigne Cedex, France\\
\{meriam.benghorbel,frederic.cuppens,nora.cuppens\}@enst-bretagne.fr\\
$^b$SUP'COM Tunis, \\ Route de Raoued Km 3.5, 2083 Ariana, Tunisie\\bouhoula@planet.tn\\
}

\maketitle
\thispagestyle{empty}

\begin{abstract}
In the field of access control, delegation is an
important aspect that is considered as a part of the
administration mechanism. Thus, a complete access
control must provide a flexible administration model to
manage delegation. Unfortunately, to our best
knowledge, there is no complete model for describing
all delegation requirements for role-based access
control. Therefore, proposed models are often
extended to consider new delegation characteristics,
which is a complex task to manage and necessitate the
redefinition of these models.

In this paper we describe a new delegation
approach for extended role-based access control
models. We show that our approach is flexible and is
sufficient to manage all delegation requirements.
\end{abstract}

\Section{Introduction}

The delegation is the process whereby a user
without any administrative prerogatives obtains the
ability to grant some authorizations. In the field of
access control, defining concept of delegation is a
difficult issue and only few works are dedicated to this
point ~\cite{rbdm, rbdm1, crampton, abdm, pbdm}.

These works showed that delegation is a complex problem to solve
and is generally modeled separately from other administration
requirements. The reason is that proposed models are generally
based on the RBAC model ~\cite{rbac} (Role-Based Access Control),
which is not expressive enough to deal with delegation
requirements such as temporary, partial, multi-step or multiple
delegation. For instance, in the RBAC model, the only way to grant
permission to a subject is by granting this permission to a given
role and then assigning this role to the subject. This is a rigid
scenario, which does not adequately answer the need of
fine-grained delegation.
\enlargethispage{3pt}

Therefore, it is necessary to extend the RBAC
model by adding new components, such as new types
of roles, permissions and/or relations. This is a
complex task to manage, and to our best knowledge,
there is no complete model for describing all
delegation requirements. Thus, delegation models
themselves are extended to consider new delegation
characteristics.

In this paper, we aim at proposing a flexible and complete
delegation approach for role based access control. Our work is
based on the OrBAC ~\cite{orbac} (Organization based Access
Control) formalism, which provides integrated framework to deal
with various security requirements including delegation
requirements.

Namely, the OrBAC model gives means to specify contextual
authorizations, which facilitate the modeling of delegation
characteristics such as temporary delegation, cascading
revocation, etc. In addition, in AdOrBAC ~\cite{adorbac}
(Administration model for OrBAC) a large number of conditions can
be expressed thanks to the use of views (in OrBAC model a view is
an access control entity used to put together objects to which
apply the same authorizations, for more details see
~\cite{adorbac}).

This provides means to specify fine-grained delegation
constraints, such as prerequisite conditions for the grantor (the
user who performs the delegation) and the grantee (the user who
receives the delegation). Therefore the administrator can restrict
the delegation by specifying delegation constraints that grantor
and grantee must satisfy.

This paper is organized as follows. In section 2 we
start with basic concepts of the OrBAC and AdOrBAC
models. In section 3 we present our delegation model.
We discuss in section 4 the complexity and the
decidability of our approach. Then related work are
given in section 5. Finally, concluding remarks are
made in section 6.

\Section{Basic concepts for OrBAC}
Before presenting our delegation model, we shall
briefly recall the main components of OrBAC.

\SubSection{OrBAC model}

The central entity in OrBAC is the entity
organization. Intuitively, an organization is any entity
that is responsible for managing a security policy.

The objective of OrBAC is to specify the security
policy at the organization level that is independently of
the implementation of this policy.

Thus, instead of modeling the policy by using the
concrete concepts of subject, action and object, the
OrBAC model suggests reasoning with the roles that
subjects, actions or objects play in the organization.

The role of a subject is simply called a role,
whereas the role of an action is called activity and the
role of an object is called view.

In OrBAC, there are eight basic sets of entities: $Org$
(a set of organizations), $S$ (a set of subjects), $A$ (a set of
actions), $O$ (a set of objects), $R$ (a set of roles), $\texttt{A}$ (a
set of activities), $V$ (a set of views) and $C$ (a set of
contexts). However, for the sake of simplicity, we
assume that the policy applies to a single organization
and thus we omit the $Org$ entity in the following.

In the following we present the basic OrBAC builtin
predicates:

\begin{itemize}

\item [-] \textbf{\textit{empower}} is a predicate over domains
$S$x$R$. If $s$ a subject and $r$ a role, then $empower(s, r)$
means that subject $s$ is empowered in role $r$.

\item [-] \textbf{\textit{use}} is a predicate over domains
$O$x$V$. If $o$ is an object and $v$ is a view, then $use(o, v)$
means that object $o$ is used in view $v$.

\item [-] \textbf{\textit{consider}} is a predicate over domains
$A$x$\texttt{A}$. If $\alpha$ is an action and $a$ is an activity,
then $consider(\alpha, a)$ means that action $\alpha$ implements
the activity $a$.

\item [-] \textbf{\textit{hold}} is a predicate over domains
$S$x$A$x$O$x$C$. If $s$ a subject, $\alpha$ an action, $o$ an
object and $c$ a context, $hold(s, \alpha, o, c)$ means that
context $c$ holds between subject $s$, action $\alpha$ and object
$o$.

\item [-] \textbf{\textit{permission}},
\textbf{\textit{prohibition}} and \textbf{\textit{obligation}} are
predicates over domains $R_s$x$A_a$x$V_o$x$C$. Where $R_s $= $R
\cup S$, $A_a$ = $\texttt{A} \cup A$ and $V_o$= $V \cup O$. More
precisely, if $g$ is a role or a subject, $t$ is a view or an
object and $p$ is an activity or an action, then $permission(g, p,
t, c)$ (resp. $prohibition(g, p, t, c)$ or $obligation(g, p, t,
c)$) means that, grantee $g$ is granted permission (resp.
prohibition or obligation) to perform privilege $p$ on target $t$
in context $c$.

\end{itemize}

Since in the OrBAC model it is possible to specify both
permissions and prohibitions, some conflicts may occur. In
prioritized OrBAC ~\cite{conflict} the authorization rules are
associated with priorities in order to evaluate their significance
in conflicting situations. Then, predicates $permission(g, p, t,
c)$ and $prohibition(g, p, t, c)$ are replaced by: $permission(g,
p, t, c, l)$ and $prohibition(g, p, t, c, l)$, where $l$ is the
priority level.

The OrBAC model is self administrated, that is the
concepts used to define an administration policy are
similar to the ones presented in this section. We give in
the following section the basic concepts of the
AdOrBAC model.

\SubSection{AdOrBAC model}

The approach in AdOrBAC is to define administration functions by
considering different administrative views. Objects belonging to
these views have specific semantics; More precisely, we shall
consider in the following two administrative views called {\em
role\_assignment} and {\em license} views. They are respectively
interpreted as an assignment of a user to a role in the {\em
role\_assignment} view and a permission to a role or to a user in
the {\em license} view.

Intuitively, inserting an object in these views will
enable an authorized user to respectively assign a user
to a role, assign a permission to a role or assign a
permission to a user. Conversely, deleting an object
from these views will enable a user to perform a
revocation.

Defining the administration functions in AdOrBAC
then corresponds to specifying which role is permitted
to have an access to administrative views. So that only
valid licenses can be created.

The AdOrBAC approach is homogeneous with the
remainder of the OrBAC model. The syntax used to
define permission to administer the policy is
completely similar to the remainder of OrBAC.

The two administrative views \textit{license} and
\textit{role\_assignment} are defined as follows:

\paragraph{License view} is used to specify and manage the
security policy. Objects belonging to the license view
have the following attributes: \textit{grantee}: subject to which
the license is granted, \textit{privilege}: action permitted by the license,
\textit{target}: object to which the license grants an access and
\textit{context}: specific conditions that must be satisfied to
use the license.

The existence of a valid license is interpreted as a
permission by the following rule:
\vspace{0.6ex}

\noindent$permission(Sub,Act,Obj,Context)Ê$:-

$use(L,license), grantee(L,Sub),$

$privilege(L,Act), target(L,Obj),$

$context(L,Context).$
\vspace{0.6ex}

\paragraph{Role\_assignment view} is associated with the following attributes: \textit{assignee}: subject to which the role
is assigned and \textit{assignment}: role assigned by the role assignment.

There is the following rule to interpret objects of the \textit{role\_assignment} view:
\vspace{0.6ex}

\noindent$empower(Subject,Role)$:-

$use(RA,role\_assignment),$

$assignee(RA,Subject), assignment(RA,Role).$

\section{Delegation model}

We define in this section our delegation model. We
show that the expressiveness of AdOrBAC is sufficient
to manage delegation requirements without adding new
components to the model.

We present in this section the main delegation
characteristics ~\cite{rbdm} like totality, permanence,
monotonicity, levels of delegation, multiple delegation,
cascading revocation, grant-dependency and show how
to model them using the OrBAC formalism.

The approach we suggest to manage delegation is the use of the notion of contexts and administrative views as defined in AdOrBAC.

We define two administrative views: the \textit{license\_delegation} view and the\textit{ role\_delegation} view. These views are used, respectively, to delegate rights (partial delegation) and roles (total delegation) and they are defined as follows:
\vspace{0.6ex}

\noindent$permission(GR, A, O, C)$ :-

$use(L, license\_delegation), grantee(L, GR),  $

$privilege(L, A), target(L, O), context(L, C). $ \vspace{0.6ex}

\noindent$empower(GR, Role)$ :-

$use(RD, role\_delegation), $

$assignee(RD, GR), assignment(RD, Role).$ \vspace{0.6ex}

Objects belonging to these views have the same attributes, respectively, as the\textit{ license} view and the \textit{role\_assignment} view but also have an additional attribute called \textit{grantor} : the subject who is creating the license or the role.

Inserting an object in the \textit{license\_delegation} view or in
the \textit{role\_delegation} view will enable a grantor to
respectively delegate permission and role to a grantee. Therefore
to manage delegation policy we must define which grantor (role or
user) have an access to these views and in which context. This is
defined by facts having the following form: 

\noindent$permission(gr, delegate, license\_delegation, context).$

\vspace{0.6ex} \noindent$permission (gr, delegate,
role\_delegation, context).$ \vspace{0.6ex}

To illustrate the approach, let us consider a situation where there are two users, \textit{John} a professor and \textit{Mary} his secretary. The role secretary is not generally permitted to have an access to the view \textit{stud\_notes}. However, \textit{John} decides to delegate to\textit{ Mary} a permission to update his student's notes. Obviously, \textit{John} must have a permission to delegate this right.

For this purpose, the administrator should first create the following administrative view:

\vspace{0.6ex}
\noindent$use(L,note\_delegation)$:-

$use(L, license\_delegation),$

$privilege(L, update), target(L, stud\_notes).$ \vspace{0.6ex}

The view \textit{note\_delegation} is derived from \textit{license\_delegation} view and only contains licenses to update student's notes.

The administrator should then give to the role professor the permission to delegate licenses in this view:
\vspace{0.6ex}

\noindent$permission(prof, delegate, note\_delegation,
nominal).$ 

\noindent where $nominal$ represents the default context.

\vspace{0.6ex} Using this permission, \textit{John} can delegate
to \textit{Mary} a permission to update his student's notes by
creating a new license $L_1$, in the \textit{note\_delegation}
view, with the following attributes: \textit{grantee}: Mary,
\textit{privilege}: update, \textit{target}: John\_stud\_notes
(which is a sub-view of \textit{stud\_notes}) and
\textit{context}: nominal. As a result, the following permission
is created:

\vspace{0.6ex} \noindent$permission(mary, update,
john\_stud\_notes, nominal).$

\vspace{0.6ex}
Notice here that John can delegate the permission to update the view  \textit{John\_stud\_notes} because this is a sub\_view of  \textit{stud\_notes}.

Therefore we assume that if the grantor have the permission to
delegate the license  \textit{L} then he also have the permission
to delegate a license $L$', which is a sub\_license of \textit{L}.

We shall now formally define different types of delegation
parameters, namely permanence, monotonicity, levels of delegation,
multiple delegation; cascading revocation and grant dependent
revocation. For this purpose, we need define the following
predicate to verify if the license \textit{L} is a sub\_license of
$L$': \vspace{0.6ex}

\noindent$sub\_license(L,L')Ê$:-

$target(L,T), target(L',T'), sub\_target(T,T'),$

$privilege(L,P),privilege(L',P'), sub\_priv(P,P'),$

$context(L,C),context(L',C'),sub\_context(C,C').$ \vspace{0.6ex}

We consider  \textit{O} is a sub\_target of  \textit{O'} if there
are two views and  \textit{O} is a sub\_view of  \textit{O'}, or
if \textit{O} is an object used in the view  \textit{O'}, or if
they are equal. \vspace{0.6ex}

\noindent$sub\_target(O,O')Ê$:-

$sub\_view(O, O'); use(O, O'); O=O'.$ \vspace{0.6ex}

Similarly, we define predicates  \textit{sub\_privilege} and  \textit{sub\_context}.

We also define the following predicate to verify if licenses  \textit{L} and  \textit{L'} are equivalent:
\vspace{0.6ex}

\noindent$equiv\_licenses(L,L')$: -

$sub\_license(L,L') , sub\_license(L',L).$

\SubSection{Permanence}

Permanence refers to types of delegation in terms of their time duration. Indeed, in some circumstances, the delegation only applies temporarily and will be automatically revoked after a given deadline. This may be modeled in our approach by simply using a temporal context. For further details about the context definition see ~\cite{context}.

In the previous example, there is no temporal specification of time duration in the context of delegation, so the delegation is permanent. Now if we assume that  \textit{John} wants to delegate to  \textit{Mary} the permission to update his student's notes only during his vacation, then he must specify this condition in the delegation context associated with the new license  $L_2$ he creates for  \textit{Mary}. The license  $L_2$ is similar to  $L_1$ except that context = during\_John\_vacation. The delegated permission is specified as follows:
\vspace{0.6ex}

\noindent$permission(mary,update, john\_stud\_notes, $

\hfill $during\_john\_vacation).$

\SubSection{Monotonicity}

Monotonic delegation means that upon delegation the grantor maintains the permission he has delegated, as described in the example of previous sections. On the other hand, with a non-monotonic delegation, the grantor loses this permission for the duration of the delegation.

To model non-monotonic delegation we define the  \textit{license\_transfer} view as follows:
\vspace{0.6ex}

\noindent$use(L,license\_delegation) $:-

$use(L, license\_transfer).$ \vspace{0.6ex}

\noindent$prohibition(Sub, Act, Obj, C, Max)$:-

$use(L ,license\_transfer), $

$grantor(L, Sub), privilege(L, Act), $

$target(L, Obj),context(L, C).$ \vspace{0.6ex}

The \textit{license\_transfer} view is a sub\_view of the  \textit{license\_delegation} view. So, inserting an
object in this view will create a new permission to the
grantee. In addition, it will create an interdiction to the
grantor associated with the highest priority level Max.
Therefore, the grantor will lose the permission he has
delegated.

Note that, the context of the prohibition and the delegated permission is the same one. So the grantor will lose this permission only for the time of the delegation.

\subsection{Multiple delegation}

Multiple delegation refers to the number of grantees to whom a
grantor can delegate the same right at any given time. To control
the delegation we assume that this number (\textit{Nm}) is fixed
by the administrator using the context
\textit{max\_multi\_delegation}. In simple delegation case
\textit{Nm} is equal to 1: \vspace{0.6ex}

\noindent$permission(subject , delegate, view, context $

\hfill $\& max\_multi\_delegation(Nm)).$ \vspace{0.6ex}

To define the context  \textit{max\_multi\_delegation} we need to count the delegation number concerning the same grantor and the same right:
\vspace{0.6ex}

\noindent$hold(S, A, L, max\_multi\_delegation(Nm)) $:-

$use(L, license\_delegation), $

$grantor(L, S), count(L', use(L', license\_delegation), $

$grantor(L',S), equiv\_licenses(L, L'),Nm'), $

$Nm'<=Nm.$ \vspace{0.6ex}

We assume that  \textit{count(V, p(V),N)} is a predicate that count the set of instances of variable  \textit{V} that satisfies predicate  \textit{p(V)}.  \textit{N} represents the result of the count predicate.

Note that, we consider the licenses  \textit{L} and  \textit{L'} are the same right since they are equivalent.

To explain this, let us consider the same roles of the previous example. Suppose now we are in a simple delegation case, so that  \textit{John} can delegate the right to update his student's notes only for one time:
\vspace{0.6ex}

\noindent$permission(john, delegate, note\_delegation, $

\hfill  $max\_multi\_delegation(1)).$ \vspace{0.6ex}

Thus, if  \textit{John} delegate to  \textit{Mary} the permission to update the view  \textit{John\_students\_notes}, then  \textit{John} does not have the permission to delegate this right to another user. For instance, he cannot delegate the permission to update the file  \textit{master\_stud\_notes}, which belongs to the view  \textit{John\_stud\_notes}, to his assistant since this right is a sub\_license of the first one.

Notice that when the  \textit{max\_multi\_delegation} context is not used, the number of permissions a subject can delegate is not restricted. So this subject can delegate as many licenses he wants.

\subsection{Level of delegation}

This characteristic defines whether or not each delegation can be further delegated and how many times.

For this purpose, we define the \textit{grant\_option\_licence} view as follows :
\vspace{0.6ex}

\noindent$permission(U, delegate, Licence, C\&valid\_level)$:-

$use(L, grant\_option\_license), grantee(L, U),$

$ target(L, Licence),context(L,C).$ \vspace{0.6ex}

The context \textit{valid\_level} is defined as follows:
\vspace{0.6ex}

\noindent$hold(U, delegate, L, valid\_level)$:-

$use(L', grant\_option\_license), sub\_licence(L', L), $

$grantor(L', U), level(L,V), level(L',V'), V' < V.$ \vspace{0.6ex}

Objects belonging to the\textit{ grant\_option\_licence} view have
an additional attribute called \textit{level}: the number of
authorized delegation steps.

Inserting a license in this view will create a permission to the grantee to delegate the right but only in the context valid\_level.

This means that, if we consider the same license  $L_1$ of the previous example and suppose that  \textit{John} wants to grant his secretary the permission to delegate this license with a delegation level equal to 3.

For this purpose, \textit{John} creates in the \textit{ grant\_option\_licence} view the licence $L_3$ with the following attributes: \textit{grantee}: Mary, \textit{privilege}: delegate, \textit{target}: $L_1$, \textit{level}: 3, \textit{context}: nominal.

This corresponds to the following rule:
\vspace{0.6ex}

\noindent$permission(mary, delegate, L_1, valid\_level).$
\vspace{0.6ex}

Therefore  \textit{Mary} can delegate the license $L_1$ (or a sub\_license of $L_1$) in the context  \textit{valid\_level}, which means that she can create a license  $L_4$ to grant another user to delegate this license and the delegation level of  $L_4$ must be lower than 3, since the delegation level of  $L_3$ is equal to 3.

The grantor can also restrict the scope of the delegation using conjunctive context. For instance,  \textit{John} can specify, in the delegation context, that  \textit{Mary} can grant another user to delegate $L_1$ only during her vacation:
\vspace{0.6ex}

\noindent$permission(mary, delegate, L1, valid\_level$

\hfill  $\& during\_mary\_vacation).$ \vspace{0.6ex}

In this case,  \textit{Mary} can further delegate the license she receives from  \textit{John} but the delegated license will only apply in the context  \textit{during\_Mary\_vacation}.

Note that we consider here only the monotonic and the partial delegation. The \textit{ grant\_option\_licence} view is also used in the total delegation case (multi-step role delegation) and non\_monotonic delegation case (multi-step transfer). A more detailed model will be proposed in future work.

\subsection{Revocation}

Revocation is an important aspect in delegation models.
In this section we present some revocation properties
and we plan to give a more detailed presentation in a
forthcoming paper.\\

\noindent \textbf{Grant Dependency  }  In the case of Grant\_Dependent revocation (GD) only the grantor is allowed to revoke the delegated license or role. On the other hand, Grant\_Independent revocation (GID) allows any member in the sponsoring role to revoke the grantee. This is modeled using contexts:

\vspace{0.6ex} \noindent$permission(subject, revoke,
license\_delegation, gd).$

\vspace{0.6ex} \noindent$permission(subject, revoke,
license\_delegation, gid).$ \vspace{0.6ex}

The  \textit{gd} and  \textit{gid} contexts are defined as
follows:

\vspace{0.6ex} \noindent$hold(User, revoke, L, gd)$:-

$use(L,  license\_delegation), grantor(L, User).$ \vspace{0.6ex}

\pagebreak
\noindent$hold(User, revoke, L, gid)Ê$:-

$use(L,  license\_delegation), grantor(L, GR),$

$empower(GR, Role), empower(User, Role).$ \vspace{0.6ex}

Contexts  \textit{gd} and  \textit{gid} are relevant for
license revocation, we can similarly define  \textit{gdr} and  \textit{gidr} to revoke a role.\\


\noindent \textbf{Cascading revocation  } In multi-step delegation it is necessary to give the possibility to revoke indirectly the delegation chain.

We can model this property thanks to the contextual license: the delegation of right is valid only if the grantor still has this right.

For this purpose we define the view  \textit{cascading\_delegation}, which is a sub\_view of  \textit{license\_delegation} view, as follows:
\vspace{0.6ex}

\noindent$permission(Sub, Act, Obj, C\& valid\_deleg(gr))$:-

 $use(L,cascading\_delegation), grantee(L,Sub), $

 $grantor(L, gr), privilege(L,Act),$

 $target(L,Obj), context(L,C).$
\vspace{0.6ex}

Inserting an object in this view will create a permission with an
additional context (\textit{valid\_deleg context}) which verify if
the grantor still has his right.

This context is defined as follows:
\vspace{0.6ex}

\noindent$hold(User, A, O, valid\_deleg(gr))Ê$:-

$is\_permitted(User, A, O).$ \vspace{0.6ex}

Therefore, the delegated permission is valid only if the delegation chain is maintained.

Note that this property only concerns monotonic delegation. Other revocation aspects remain to be investigated in further work.

\section{Decidability and complexity}

The OrBAC model is based on first order logic and more precisely
on Datalog ~\cite{datalog} which ensures a decidable and tractable
theory.

Datalog programs do not allow the use of functional terms and must
only include both defined and safe rules. A rule is defined if
every variable that appears in the conclusion also appears in the
premise. A rule is safe if it only provides means to derive a
finite set of new facts. In pure Datalog program, rules do not
contain any negative literal. Pure Datalog guarantees that any
access control policy will be decidable in polynomial time.
However pure Datalog expressivity is very restricted.

In Datalog$^\neg$, the negation restriction is relaxed. Negative
literals are allowed but rules must be stratified ~\cite{datalog}.
A stratified Datalog$^\neg$ program is computable in polynomial
time.

The definition of security policies using the OrBAC model obeys
the Datalog$^\neg$ restriction except the definition of contexts
through the  \textit{hold} predicate. More precisely, the security
rules correspond to ground close facts specified using the
\textit{permission},  \textit{prohibition} predicates.
Specifications of predicates  \textit{empower},  \textit{use} and
\textit{consider} correspond also to facts or rules that respect
the Datalog$^\neg$ restriction.

By contrast, the definition of contexts does not correspond to
Datalog$^\neg$ restriction for the following reasons: these rules
contain functional terms and are not always safe and defined.

To solve these problems it is proposed firstly, to restrict the
theory so that only  \textit{relevant} contexts are evaluated. A
context is relevant if it appears in the definition of a security
rule. Secondly, a relevant context is always fully instantiated.
Finally it is proposed to pre-compute the evaluation of the
\textit{Empower},  \textit{Use} and  \textit{Consider} predicates
using a bottom-up strategy. Then, the evaluation of queries is
completed using the top-down strategy as defined in the SGL
algorithm ~\cite{toman}. This hybrid strategy guarantees the
decidability of query evaluation in the OrBAC model and its
termination in polynomial time.

\section{Related work}

The previous work on delegation has shown that delegation is a
complex concept and, to our best knowledge, there is no complete
model for describing all delegation characteristics such as
multiple delegation, cascading revocation, etc.

Proposed models are based on RBAC which is not expressive enough to deal with the delegation requirements. To solve this problem it is suggested to extend the RBAC model to include delegation components, such as new types of roles, actions, permissions, etc. Unfortunately, this is a complex task to manage, since it is necessary to add new components for modeling every delegation characteristic.

For instance, in the RBDM0 model proposed by ~\cite{rbdm}, authors extend RBAC$_0$ model to define role-based delegation. They define a relation can-delegate $\subseteq R$x$R$ to control role delegation and add new components such as: Users-O and Users-D to differentiate between original and delegated members, UAO and UAD to specify original member assignment and delegate member assignment relations, etc.

They also propose some extensions to RBDM0 to address more delegations characteristics. This requires additional components. For instance they add new types of permissions: delegable and non delegable permissions (permissions-PN and permissions-PD) to model partial delegation.

In RBDM1 ~\cite{rbdm1}, an extension of RBDM0, is proposed. This model adds new components such as a partially ordered role hierarchy relation $RH \subseteq R$ x $R$ to model delegation using hierarchical roles.

The PBDM model ~\cite{pbdm} is another delegation model based on RBAC96. This model uses the can-delegate relation with prerequisite condition to restrict delegates, and adds new types of roles and permissions to address permission level delegation requirement. In PBDM0 roles are partitioned into regular roles (RR) and delegation roles (DTR). This partition induces a parallel partition of the two RBAC components: user-role assignment (UA) and permission-role assignment (PA). UA is separated into user-regular role assignment (UAR) and user-delegation role assignment (UAD). PA is similarly separated into permission-regular role assignment (PAR) and permission-delegation role assignment (PAD).

PBDM1 is an extension to PBDM0 which supports security administrator involved delegation and revocation. This model adds new components such as delegatable roles (DBR), user-delegatable role assignment (UAB) and permission-delegatable role assignment (PAB).

PBDM2 model is another extension which addresses a role-to-role delegation. Like other models, PBDM2 adds new components, such as temporal delegatable roles (TDBR), and redefines existing ones.

The ABDM model ~\cite{abdm} is an attribute-based delegation model, which extend PBDM model to address delegation constraint. This model redefines the can-delegate relation to restrict the delegation. The ABDM model is also extended to ABDM$_x$ for more flexibility.

Another delegation model is proposed in ~\cite{crampton}. This model is more complete than previous works. But it also extends RBAC96 by adding new components to specify more delegation characteristics like temporary non-monotonic delegation. It introduces the relations: can-delegate and can-receive to authorize role delegation, and the relations can-delegatep and can-receivep to authorize permission delegation. Also, it defines actions like xferR$_0$, xferP$_0$, xferP$_1$, etc, to model roles and permissions transfer. The two relations tempUA and tempPA are introduced to record temporary user-role and user-permission delegations.

Like the above discussed works, this model introduces new relations or actions to model each delegation characteristics. This is a complex task to manage especially when the delegation model has to be enriched.

Compared to these works, our model is more flexible, simpler to manage and more complete. Indeed, OrBAC model offers facilities to deal with delegation requirements without the need for additional components.

Namely, OrBAC model is based on multi-granular and contextual licenses. This provides facilities to define many delegation characteristics like totality, permanence, revocation, etc.

Moreover, thanks to the use of views we can express a large number
of conditions, which allow us to specify delegation constraints;
This is modeled by prerequisite conditions associated with the
grantee or the grantor. For instance, the professor is permitted
to delegate a permission to manage her courses to her assistant
but only if this assistant is a graduate student.

\section{Conclusion}

In this paper, we have proposed a new delegation approach for role-based access control. We have showed that it is possible to specify delegation requirements using the OrBAC formalism. This model is self administrated and offers facilities, such as multi-granular license, contextual license, use of views, etc., which gives means to specify delegation characteristics without adding new components or modifying the exiting ones. Therefore our approach is more flexible, simpler and more complete than previous works based on RBAC model.

The future work will be dedicated to enrich our delegation model and more precisely the revocation mechanism. We intend to include several of the revocation schemes as described in ~\cite{revocation}.

\section*{Acknowledgment}

This work is partially supported by the RNRT project Politess. For
this work, Meriam Ben Ghorbel-Talbi is funded by the IFC, the French Institute for Cooperation in Tunisia.


\enlargethispage{-197pt}

\bibliographystyle{latex8}

\bibliography{bib}

\end{document}